Wireless data transmission in a 560-GHz band using low-phase-noise terahertz wave generated by photomixing of a pair of distributed feedback lasers injection-locking to Kerr micro-resonator soliton comb

Yu Tokizane,[1,†,*] Shota Okada,[2] Takumi Kikuhara,[2] Hiroki Kishikawa,[1] Yasuhiro Okamura,[1,3] Yoshihiro Makimoto[1,4], Kenji Nishimoto[1],Takeo Minamikawa,[1] Eiji Hase,[1] Jun-Ichi Fujikata,[1] Masanobu Haraguchi,[1] Atsushi Kanno,[5,6] Shintaro Hisatake,[7] Naoya Kuse,[1] and Takeshi Yasui[1,†,*]

[1]Institute of Post-LED Photonics (pLED), Tokushima University, 2-1, Minami-Josanjima, Tokushima 770-8506, Japan

[2]Graduate School of Sciences and Technology for Innovation, Tokushima University, 2-1, Minami-Josanjima, Tokushima 770-8506, Japan

[3]Center for Higher Education and Digital Transformation, University of Yamanashi, 4-4-37 Takeda, Kofu, Yamanashi 400-8510, Japan

[4]Tokushima Prefectural Industrial Technology Center, Saika-cho, Tokushima, Tokushima 770-8021, Japan

[5]National Institute of Information and Communications Technology, 4-2-1 Nukuikitamachi, Koganei, Tokyo 184-8795, Japan

[6]Department of Electrical and Mechanical Engineering, Nagoya Institute of Technology, Gokiso-cho, Showa-ku, Nagoya, Aichi 466-8555, Japan

[7]Electrical and Energy System Engineering Division, Gifu University, 1-1 Yanagido,

Gifu 501-1193, Japan

†Authors contributed equally as corresponding authors.

*tokizane@tokushima-u.ac.jp,　*yasui.takeshi@tokushima-u.ac.jp

**Abstract**

The demand for higher data rates in next-generation mobile wireless communication systems (6G) has led to significant interest in terahertz (THz) waves as a high-frequency, broad modulation bandwidth carrier wave. In this study, we propose and demonstrate a wireless data transfer in the 560-GHz band using low-phase-noise THz waves generated by photomixing of a pair of distributed feedback lasers injection-locking to Kerr micro-resonator soliton comb. Experimental results showed near-error-free on-off keying (OOK) data transfer at 1 Gbit/s in the 560-GHz band, with a Q-factor of 6.23, surpassing the error-free limit. Also, modulation formats of binary phase shift keying (BPSK) and quadrature phase shift keying (QPSK) were successfully used, showing clear constellation diagrams and relatively low root mean squared error vector magnitude (rms EVM) values of 23.9% and 23.6%, respectively. Moreover, data transfer at 0.4 Gbit/s in 16 quadrature amplitude modulation (16QAM) demonstrated clear isolated symbols and achieved a low rms EVM value of 8.1%, complying with the IEEE 802.15.3d standard amendment. These demonstrations highlight the potential of using injection-locked DFB lasers with the Kerr micro-resonator soliton comb to achieve high-quality, high-speed wireless data transfer in the 560-GHz band. These findings contribute significantly to the advancement of wireless communication technology in the THz frequency range and pave the way for the realization of 6G wireless communication systems.

## 1. Introduction

Terahertz (THz) wave is expected for a carrier wave in next-generation mobile wireless communication (6G, expected carrier frequency > 300 GHz) because it can be used as a high-frequency, broad modulation bandwidth, carrier wave to largely accelerate the data rate over present wireless communications (5G, carrier frequency = 28 GHz or more) [1]. In particular, broadband data transmission in frequencies higher than 350 GHz has attracted attention for wireless mobile fronthaul and backhaul in 6G to provide a higher data rate while avoiding interference with general 6G mobile communication and other wireless applications. The bottleneck in making full use of 6G is the performance limit of THz emitters. Electronic THz emitters based on frequency multiplication of a master clock signal have been successfully used until 5G; however, it may face technical barriers in 6G because 6G carrier frequency is too high for wireless electronics. One of technical barriers is increased phase noise of THz wave arising from higher-order frequency multiplication.

Use of photonic technology in generation of THz wave can suppress the increased phase noise of THz wave. In photonic THz generation, one of the key elements expected to play a crucial role is the optical frequency comb (OFC) such as mode-locked fiber-laser OFC [2, 3] and electro-optic-modulator OFC [4,5] because its frequency spacing $f_{rep}$ is inherently stable and its relative phase noise is considerably low owing to mode-locking oscillation and active control. Photomixing of internal OFC modes enable the generation of stable THz wave at a harmonic band of $f_{rep}$. For example, a uni-travelling carrier photodiode (UTC-PD) [6, 7] generates the THz wave from two OFC modes via photo-mixing process without spoiling the low phase noise of $f_{rep}$ inherent in OFC. Therefore, a combination of mode-locked fiber-laser OFC (ML-FL-OFC) or electro-optic-modulator OFC (EOM-OFC) with UTC-PD is a promising way to generate low-phase-noise microwave and millimeter wave [2-5]. However, when this approach is extended to THz waves, two OFC modes with a large frequency separation (= $mf_{rep}$) must be used for photomixing because the THz wave frequency $f_{THz}$ is much larger than the $f_{rep}$ of those OFCs (typically, 100 MHz for ML-FL-OFC and 10 GHz for EOM-OFC). Photomixing of two largely-separated OFC modes suffers from the quadratic-dependent increase in phase noise on the number of optical frequency multiplications (= $m$). If two adjacent OFC modes can be used for photomixing-based THz-wave generation in place of two largely-separated OFC modes, phase noise of THz wave should be equal to that of $f_{rep}$ in OFC.

Recently, the on-chip Kerr micro-resonator soliton comb, namely the soliton microcomb, has appeared as a small, simple, and cost-effective OFC [8, 9]. Importantly, the extremely small size of the micro-resonator largely increases $f_{rep}$ of the soliton microcomb up to 6G carrier frequency over 300 GHz. Therefore, two adjacent OFC modes extracted from the soliton microcomb can be directly used for photomixing without the need for optical frequency multiplication, and the resulting THz wave fully benefits from the low phase noise property of $f_{rep}$ in the soliton microcomb. Based on this idea, an ultralow-phase-noise THz wave was generated at 300 GHz [10-12] and 560 GHz [13] by a combination of a stabilized soliton microcomb with UTC-PD. Furthermore, focusing on the frequency band over 350 GHz available for wireless mobile fronthaul and backhaul in 6G, on-off keying (OOK) wireless data transmission (data rate = 2 Gbit/s) was demonstrated in 560-GHz band by using a combination of a free-running soliton microcomb with UTC-PD for THz generation and a Schottky barrier diode (SBD) for THz detection [14]. However, the Q-factor of the eye pattern remained at 3.40. Although this Q-factor satisfied the limit of forward error correction (FEC limit, Q-factor = 2.33), it should be further improved for high-quality data transmission. One reason for the relatively low Q-factor is the limited optical signa-to-noise ratio (OSNR) owing to co-existence of two extracted microcomb modes and amplified spontaneous emission (ASE) background resulting from a high-gain fiber optical amplification of low-power soliton microcomb. Optical injection locking (OIL) of a high-power CW laser to a soliton microcomb is a promising approach to transfer the phase noise of the microcomb and amplify the microcomb mode without degrading the OSNR because of eliminated ASE background [15]. Although a combination of such the OIL with wireless data transmission was demonstrated in the frequency band of 300 GHz [16], there is no attempts in the frequency band over 350 GHz.

In this article, we first injection-lock a pair of distributed feedback laser diodes (DFBs) to two adjacent modes of a 560-GHz-spacing soliton microcomb. The resulting low-phase-noise, high-OSNR, DFB output lights were used for THz generation via photomixing with UTC-PD. Then, we demonstrate near error-free OOK wireless data transmission in 560-GHz by using SBD for THz detection. We further combined this system with binary phase shift keying (BPSK), quadrature phase shift keying (QPSK), and 16 quadrature amplitude modulation (16QAM) to evaluate the signal quality in advanced modulation format.

## 2. Experimental setup

Figure 1 shows a schematic of experimental setup. We generated a 560-GHz-spacing soliton microcomb (center wavelength = 1560.7 nm, power = 1 mW) from a $Si_3N_4$ (SiN) ring-shaped micro-resonator (custom, LIGENTEC, S.A., free spectral range = 560 GHz, Q factor ≈ $10^6$). The detail of the soliton microcomb is given elsewhere [8, 9]. After splitting with a fiber coupler, a half of the microcomb passed through a tunable ultra-narrowband optical bandpass filter (OBPF1; Alnair labs, TFF-15-1-PM-L-100-FA, passband wavelength = 1520 ~ 1580 nm) to extract a microcomb mode at 1551.6 nm (µOFC-M1, power = 1 µW) whereas the other half passed through another tunable ultra-narrowband optical bandpass filter (OBPF2; Alnair labs, TFF-15-1-PM-L-100-FA, passband wavelength = 1520 ~ 1580 nm) to extract a microcomb mode at 1547.2 nm (µOFC-M2, power = 1 µW). The extracted µOFC-M1 and µOFC-M2 were used for a pair of master lasers in OIL. We used a pair of 100-mW distributed feedback laser diodes (DFB1, Gooch & Housego, AA1408-193200-100-PM250-FCA-NA, $\nu_1$ = 193.2 THz corresponding to a wavelength of 1552.8 nm; DFB2, Gooch &

Housego, AA1408-193700-100-PM900-FCA-NA, $\nu_2$ = 193.7 THz corresponding to a wavelength of 1548.7 nm) for a pair of slave lasers. The μOFC-M1 and μOFC-M2 were respectively incident into DFB1 and DFB2 via a pair of optical circulators. When $\nu_1$ and $\nu_2$ were respectively tuned to the optical frequencies of μOFC-M1 and μOFC-M2 within the locking range of DFB1 and DFB2 (typically a few hundred MHz) by their current control, their OIL was achieved. The details of OIL to the soliton microcomb are given elsewhere [15]. The DFB2 light injection-locking to μOFC-M1 was delivered to a LiNbO$_3$ intensity modulator (LN-MOD, T.MXH1.5-20PD-ADC-LV, Sumitomo Osaka Cement Co., Ltd, wavelength = 1.55 μm, modulation rate = 40 Gbit/s, optical bandwidth > 20 GHz). The modulation signal was generated depending on the modulation format by an arbitrary waveform generator (AWG, M8196A, Keysight Tech., Inc., sample rates < 92 GS/s, analog bandwidth = 32 GHz), and was then applied to the LN-MOD. We used a 1-Gbit/s non-return-to-zero (NRZ) signal with a DC bais for OOK modulation signal. In BPSK, QPSK, and 16QAM, we applied the double sideband modulation at ± 5-GHz shifts from the central frequency of OIL-DFB1 as shown in the inset of Figure 1.. The modulated DFB1 light and the unmodulated DFB2 light were combined by a fiber coupler. After optical amplification with an erbium-doper fiber amplifier, the combined DFB1 and DFB2 lights (total power = 30 mW) are fed into an antenna-integrated UTC-PD (IOD-PMAN-13001-2; NTT Electronics Corp., output frequency of 300–3000 GHz, and output power of 3 μW at 600 GHz) to generate THz waves via photomixing. The THz wave propagated in free space (propagation length

= 10 mm) and then was detected by SBD (Virginia Diodes, Inc., WR1.9ZBD, RF frequency = 400–600 GHz, responsivity = 1000 V/W). The electric output signal from the SBD was amplified using an electric amplifiers (AMP, R00M30GSA, RF-LAMBDA, frequency range = 0.04–30 GHz, gain = 17 dB at 6 GHz, and SHF S824A, SHF Communication Technologies AG, frequency range = 80 kHz- 35 GHz, gain = 25 dB) via a bias-T, and was then measured by a real-time oscilloscope (UXR0402AP, Keysight Tech., Inc., max. bandwidth = 40 GHz and max. sample rate = 256 GSa/s). In BPSK, QPSK, and 16QAM, the real-time oscilloscope demodulated the signal by taking the existence of IF into account.

## 3. Result

We first evaluated the basic performance of DFB1 injection-locking to µOFC-M1 (namely, OIL-DFB1) and DFB2 injection-locking to µOFC-M2 (namely, OIL-DFB2). Figure 2(a) shows an optical spectrum of the soliton microcomb before splitting with a fiber coupler, in which multiple OFC modes with a frequency spacing of 560 GHz was confirmed in the 1550-nm band. Figure 2(b) compares the optical spectrum of µOFC (red plot) with those of OIL-DFB1 and OIL-DFB2 (blue plot). The wavelength of the OIL-DFB1 (OIL-DFB2) was the same as that of the extracted µOFC-M1(µOFC-M2), indicating successful OIL. Whereas the OSNR of the extracted µOFC-M1 and µOFC-M2 remained around 35 dB, the OIL-DFB1 and OIL-DFB2 achieve the OSNR of 53 dB. In this way, OSNR was largely enhanced without additional ASE background while transferring the low phase noise of soliton microcomb to DFB1 and OFB2. Such high-OSNR, soliton-microcomb-traceable, dual-wavelength laser light was used for generation of stable THz wave in 560-GHz band.

We next performed the OOK data transmission of a 1-GBaud baseband signal using 560-GHz THz wave as a wireless carrier. Figure 3(a) shows an eye diagram of OOK measured by the real-time oscilloscope, indicating the fully opened eye. The corresponding Q-factor was determined to be 6.23, which is close to the limit of error-free data transmission (error-free limit, Q-factor = 6.36). In this way, we achieved near-error-free OOK data transmission of 1 Gbit/s in a 560-GHz band. To investigate the effectiveness of OIL to the soliton microcomb in OOK data transmission, we also conducted the data transmission without OIL. We performed the similar experiment using the 560-GHz carrier wave generated by photomixing of two adjacent modes extracted from the 560-GHz-spacing soliton microcomb in place of OIL-DFB1 and OIL-DFB2 as demonstrated in the previous paper [14]. Figure 3(b) shows an eye diagram of OOK baseband signal of 1 GBaud in a 560-GHz band without OIL. Although the eye was opened again, the corresponding Q-factor was remained at 5.23. This Q-factor was lower than that with OIL. In this way, we confirmed the effectiveness of OIL in OOK data taransfer.

To evaluate the quality of data transmission in advanced modulation using phase and amplitude, we conducted BPSK and QPSK data transmission in a 560-GHz band with intermediate frequency of 5 GHz. Figures 4(a) and 4(b) shows constellation diagrams of BPSK and QPSK at 1 GBaud, in which red and blue show a symbol and their transition. Isolated symbols were clearly confirmed in both results. The root mean squared error vector magnitude (rms EVM) values were determined to be 23.9% for BPSK and 23.6% for QPSK. According to the literature [17], the bit-error rate (BER) could be roughly estimated to be $10^{-9}$ and $10^{-5}$ from the EVM of BPSK and QPSK, respectively, under the assumption of additive white Gaussian noise was dominant. Both cases could become error-free with the help of standard FEC technology since the BER values were below FEC limit.

We further evaluated the quality of data transmission in 16QAM data transmission in a 560-GHz band with intermediate frequency of 5 GHz. Figure 5 shows constellation diagrams of 16QAM at (a) 1 GBaud and (b) 0.1 GBaud. Unfortunately, symbols of 16QAM overlapped with each other at 1 GBaud, indicating the failed data transmission. However, at 0.1 GBaud, 16 isolated symbols were clearly confirmed. The corresponding EVM was 8.07 %, which is lower than the EVM limit of 15.8% by the IEEE 802.15.3d standard amendment [18]. The difference of constellation diagrams between them is mainly due to the difference of signa-to-noise ratio between them.

## 4. Discussion

We discuss the difference of Q factor in OOK data transmission with and without OIL. One possible reason for this difference is the presence or absence of ASE background in optical spectrum of two modes for photomixing. Figures 6(a) shows optical spectra of OIL-DFB1 and OIL-DFB2. Due to the OIL, ASE-free OSNR was achieved to 53 dB. Conversely, Fig. 6(b) shows optical spectra of two adjacent OFC modes (µOFC-M1 and µOFC-M2) extracted from the soliton microcomb. Although most of the broad ASE spectrum was eliminated by OBPF, the remaining ASE present in the extracted two-mode neighborhood is causing a decrease in OSNR. The resulting OSNR was remained around 40 dB. Since µOFC-M1 and µOFC-M2 together with the remaining ASE background could be used for photomixing, the resulting THz wave reduces its SNR and/or increases its phase noise. It is possible that this effect is leading to differences in the Q-factor depending on the presence or absence of OIL.

      We next discuss the effectiveness of the OIL in QPSK. Figure 7(a) shows the constellation diagram of the demodulated QPSK signal when the 560-GHz THz wave was generated by photomixing of two independent DFB lasers with UTC-PD. We here set the symbol rate and IF frequency to be 0.1 GBaud and 4 GHz, respectively. Sampled symbols indicated by red color indicated circular shape. Conversely, Fig. 7(b) shows the constellation diagram of the demodulated QPSK signal at the same symbol rate and IF when the 560-GHz THz wave was generated by photomixing of OIL-DFB1 and OIL-DFB2 instead of two independent DFB lasers with UTC-PD. In this case, the shape of sampled symbol indicated elliptic shape rather than the circular shape and its long axis directed towards the origin. Difference of shape between them implies the reduced phase noise thanks to the OIL.

We next discuss the effectiveness of the OIL in QPSK. Figure 7(a) shows the constellation diagram of the demodulated QPSK signal when the 560-GHz THz wave was generated by photomixing of two independent DFB lasers with UTC-PD. We here set the symbol rate and IF to be 0.1 Gbaud and 4 GHz, respectively. Sampled symbols indicated by red color indicated circular shape. Conversely, Fig. 7(b) shows the constellation diagram of the demodulated QPSK signal at the same symbol rate and IF when the 560-GHz THz wave was generated by photomixing of OIL-DFB1 and OIL-DFB2 instead of two independent DFB lasers with UTC-PD. In this case, the shape of sampled symbol indicated elliptic shape rather than the circular shape and its long axis was inclined towards the origin. Difference of shape between them implies the reduced phase noise thanks to the OIL.

Finally, we discuss a possibility to further improve the quality of wireless data transfer. A straightforward way in the side of THz generation is to increase the SNR of data transfer signal by advanced modulation. Baseband modulation in optical frequency is appropriate to increase the THz wave power because it effectively uses the optical power within the threshold of optical input power in UTC-PD. The resulting appropriate power increases conversion efficiency of UTC-PD and hence increase power of THz wave, leading to the improved SNR. Such baseband modulation also enables efficient utilization of frequency bandwidth, thereby allowing for higher data rates. In the side of THz detection, use of a mixer with a local oscillator (LO) instead of an SBD detector can enhance the power of the detected signal in the mixing process and also enhance the SNR thanks to the low noise characteristics.

## 5. Conclusion

We demonstrated wireless data transfer in the 560-GHz band using a low-phase-noise THz wave generated by photomixing OIL-DFB1 and OIL-DFB2 with UTC-PD. The injection locking of DFB lasers to adjacent modes of the soliton microcomb improved the OSNR and suppressed ASE background, resulting in a low-phase-noise THz carrier wave suitable for data transfer. In the case of OOK data transfer at 1 Gbit/s, the system achieved near-error-free performance with a Q-factor of 6.23. The introduction of OIL improved the Q-factor compared to the case without OIL. Furthermore, data transfer using more advanced modulation formats like BPSK and QPSK at 1 Gbaud showed successful constellation diagrams, indicating effective data transfer with relatively low rms EVM values. Although attempts to achieve 16QAM data transfer at 1 GBaud failed due to overlapping symbols, a reduced modulation freqecy of 0.1 Gbaud demonstrated clear isolated symbols and achieved a low rms EVM value within the limits set by standard amendments.

The study concludes that the proposed approach of using injection-locked DFB lasers with the Kerr micro-resonator soliton comb shows promise for achieving high-quality, high-speed wireless data transfer in the 560-GHz band. Further improvements can be explored by increasing the SNR through advanced modulation and employing heterodyned mixers for THz detection. These findings contribute to the advancement of wireless communication technology in the THz frequency range, which is essential for 6G.

# Figures

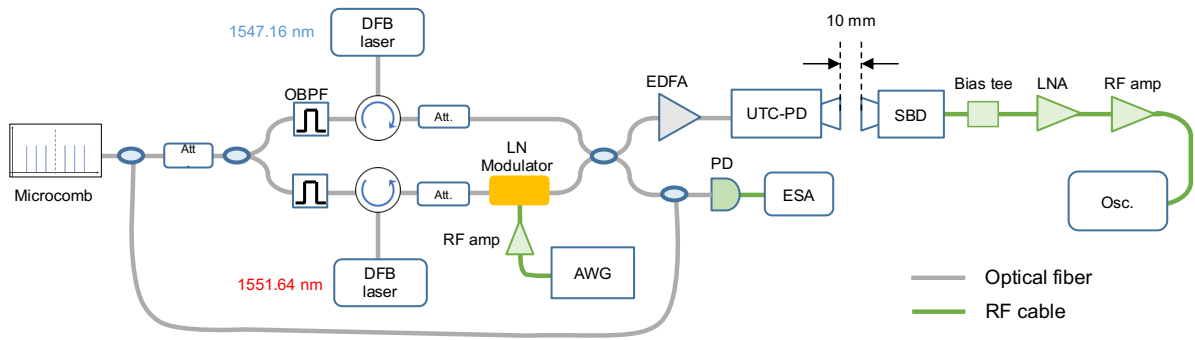

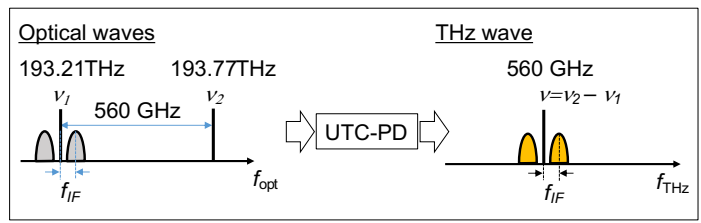

Figure 1 Experimental setup. Att., attenuator; OPBF, optical band pass filter; AWG, arbitral wave form generator; RF amp, RF amplifier; PD, photo detector; ESA, electric signal analyzer; EDFA, erbium doped fiber amplifier; UTC-PD, uni-travelling carrier photo diode; SBD, Schottky barrier diode, LNA, low noise amplifier; OSC, oscilloscope. OIL is monitored by Osc. Which measures RF spectrum of optical beat between DFB laser and microcomb mode. When the DFB is locked, beat signal disappeares.

Inset shows the schematic of optical and THz spectrum in modulation format of *M*-PSK or *M*-QAM with intermediate frequency of $f_{IF}$.

- 16 -

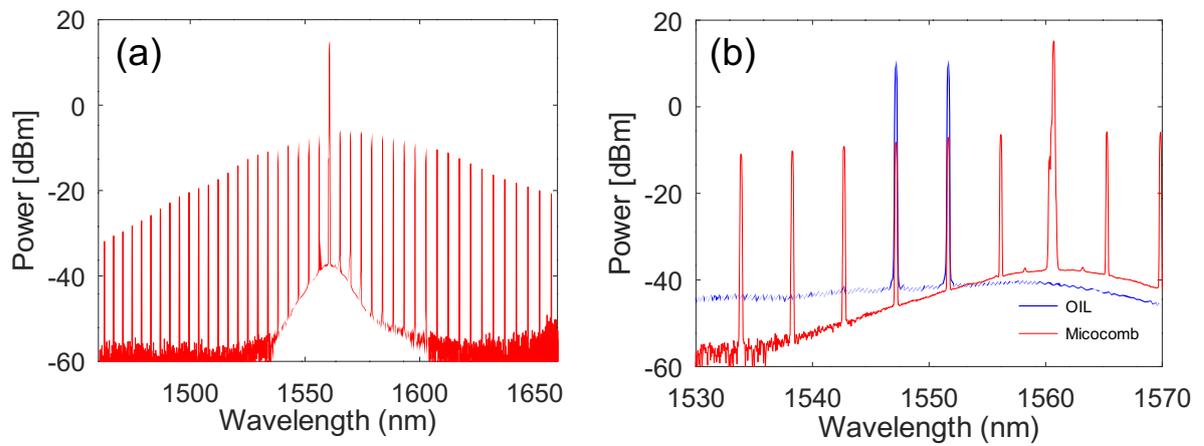

Figure 2 (a) Spectrum of soliton microcomb before optical filtering. A peaky spectrum at 1560.68 nm is sum of residual pump beam and microcomb mode. (b) Spectra of two optical modes (blue), which are optically injection locked to two microcomb modes (red). Frequency resolution of optical spectrum analyzer is 0.2 nm.

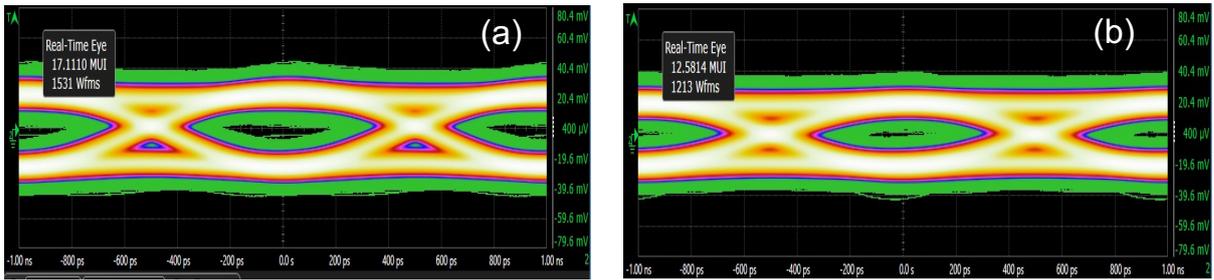

Figure 3 Eye pattern diagrams of 560 GHz OOK data transmission in modulation frequency of 1 Gbaud using (a) THz wave generated from two modes OI-Locked to microcomb modes and (b) THz wave generated from two microcomb modes filtered by optical filter. The Q-value of diagram are (a) 6.23 and (b) 5.23. Corresponding BER are $2.3\times10^{-10}$ and $8.5\times10^{-8}$.

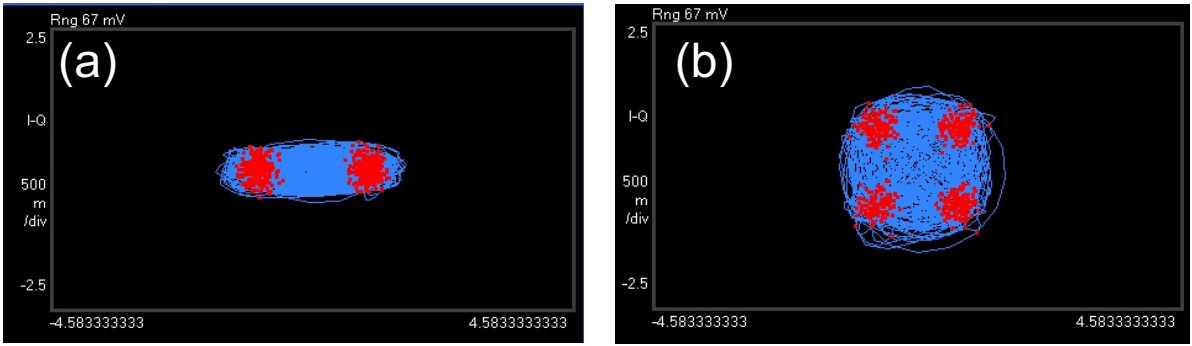

Figure 4 Constellation diagram of 560 GHz data transmission in *M*-PSK and *M*-QAM modulation format with intermediate frequency of 5GHz. (a) BPSK modulation in 1 Gbaud. (b) QPSK modulation in 1 GBaud.

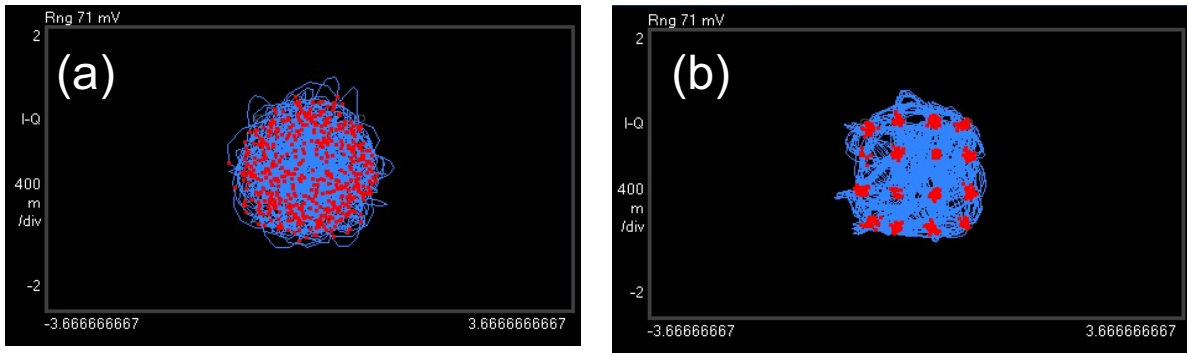

Figure 5 Constellation diagram of 560 GHz data transmission in 16 QAM modulation format with Intermediate frequency is 5GHz. (a) 16 QAM modulation in 1 GBaud. (b) 16 QAM modulation in 0.1 GBaud.

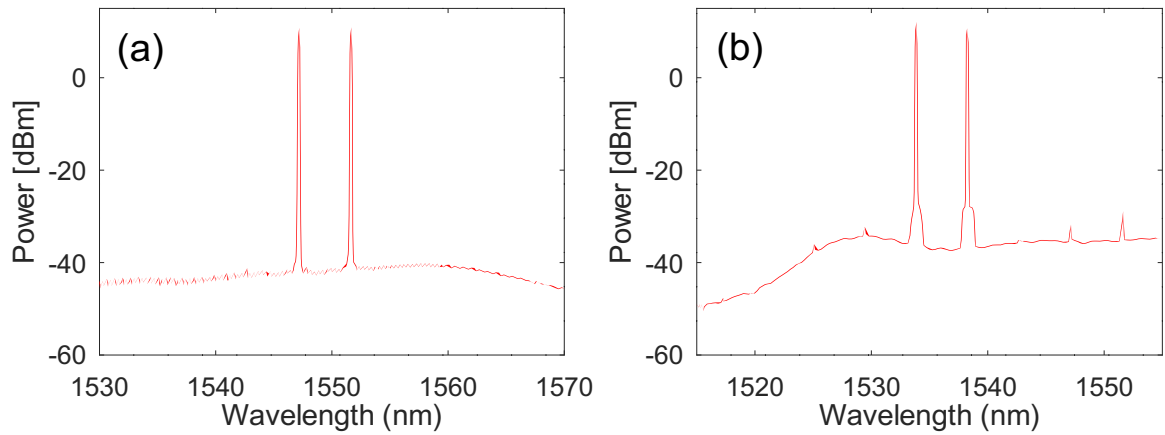

Figure 6 Comparison of optical spectra of optical modes with OIL or without OIL. (a) Two modes OI-Locked to microcomb modes. (b) Two microcomb modes filtered by optical filter. Both spectra were measured after amplification by EDFA, which are independently injected to the UTC-PD for the generation of THz wave.

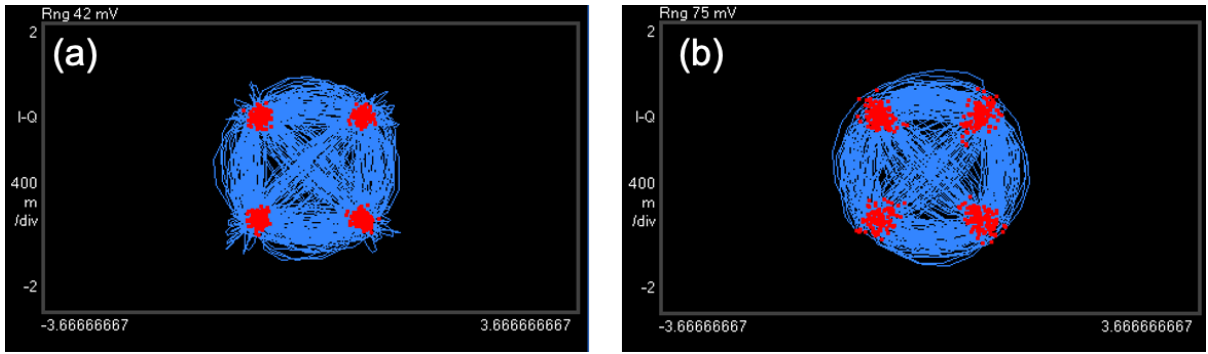

Figure 7 Comparison of constellation patterns of 560 GHz data transmission (a) with photomixing of two DFB lasers or (b) using photomixing of lights OIL to microcomb modes. The modulation format is QPSK with 0.1GBaud and intermediate frequency was 4 GHz.